\documentclass{PoS}

\usepackage{epsfig}
\usepackage{latexsym}
\usepackage{graphicx}
\usepackage{bm}
\usepackage{longtable}

\def\lsim{\raise0.3ex\hbox{$<$\kern-0.75em\raise-1.1ex\hbox{$\sim$}}}
\def\gsim{\raise0.3ex\hbox{$>$\kern-0.75em\raise-1.1ex\hbox{$\sim$}}}
\def\simgt{\rlap{\lower 3.5 pt\hbox{$\mathchar \sim$}}\raise 1.0pt \hbox {$>$}}
\def\simlt{\rlap{\lower 3.5 pt\hbox{$\mathchar \sim$}}\raise 1.0pt \hbox {$<$}}

\title{Many flavor QCD with $N_f=12$ and $16$}
\ShortTitle{Many flavor QCD with $N_f=12$ and $16$}

\author{
\speaker{Yasumichi Aoki}$^a$, Tatsumi Aoyama$^a$, Masafumi Kurachi$^a$,
Toshihide Maskawa$^a$, Kei-ichi Nagai$^a$, Hiroshi Ohki$^a$, 
Akihiro Shibata$^b$,  Koichi Yamawaki$^a$ and 

Takeshi Yamazaki$^a$

\hspace*{55mm} LatKMI Collaboration
\\ \\
$^a$
Kobayashi-Maskawa Institute for the Origin
of Particles and the Universe (KMI), Nagoya University, Nagoya
464-8602, Japan\\
$^b$
Computing Research Center, High Energy Accelerator Research Organization (KEK), 
Tsukuba 305-0801, Japan

\email{yaoki@kmi.nagoya-u.ac.jp}
}

\abstract{
Information of the phase structure of many flavor SU(3) gauge theory is of 
great interest for finding a theory which dynamically breaks the electro-weak 
symmetry. We study the SU(3) gauge theory with fermions for $N_f=12$ and 16 
in fundamental representation. 
Both of them, through perturbation theory, reside in the conformal phase. 
We try to determine the phase of each theory non-perturbatively with lattice 
simulation and to find the characteristic behavior of the physical quantities 
in the phase. HISQ type staggered fermions are used to reduce the 
discretization error which could compromise the behavior of the physical 
quantity to determine the phase structure at non-zero lattice spacings. 
Spectral quantities such as bound state masses of meson channel and meson 
decay constants are investigated with careful finite volume analysis.
Our data favor the conformal over chiral symmetry breaking scenario for
both $N_f=12$ and $16$.
}

\FullConference{The XXIX International Symposium on Lattice Field Theory - Lattice 2011\\
July 10-16, 2011\\
Squaw Valley, Lake Tahoe, California}

\begin{document}

\section{Introduction}
\label{sec:introduction}

The technicolor model 
is one of attractive candidates of the physics beyond the standard model for 
the dynamical origin of the electroweak symmetry breaking. 
In particular the gauge theory with walking behavior near the edge of the 
conformal window has been considered as a realistic new physics model 
(walking technicolor). 
Therefore it is important to understand the phase structure and the 
low-energy dynamics of these theories 
for the phenomenology to be tested by the on-going LHC experiment.
A concrete example of such models could be
the SU(3) gauge theory with 12-flavor of massless 
fermions in the fundamental representation.
The main purpose of this study is to understand the dynamics 
of this gauge theory by using the technique of the numerical 
simulation developed in lattice QCD, and 
ultimately to test whether this theory
is a candidate of the walking technicolor (see \cite{Nagai:lat2011} 
for more discussion and references).

A number of lattice works studied
this theory, such as the ones investigating the phase diagram
of Wilson fermions
\cite{{Iwasaki:1991mr},{Iwasaki:2003de}},
hadron spectrum~\cite{Jin:2009mc,Fodor:2011tu,Deuzeman:2012pv},
and the running coupling constant~\cite{{Appelquist:2007hu},{Appelquist:2009ty},{Hasenfratz:2011xn},{Aoyama:2011ry}}.
However, whether this theory is in conformal or chiral broken phase is
controversial at present.

We investigate the 12-flavor SU(3) 
gauge theory using a variant of the highly improved staggered quark (HISQ)
action \cite{Follana:2006rc}, eventually with multiple lattice spacings,
aiming to shed light on the controversy.
We present preliminary results of 
pseudoscalar meson mass and decay constant,
through which we attempt to determine the phase of this theory,
as well as the characteristic quantity associated with the phase.

Along this line we also simulate the 16-flavor SU(3) gauge theory with
the same improved action. This theory resides deep in conformal window 
according to the perturbation theory. If this is the case, this theory
could be used as a reference for the signal of conformality.
Preliminary results for this study is presented here.

Another series of simulations with this set up for $N_f<12$
have been performed to test the walking behavior of $N_f=8$,
and to have a reference for the signal of the real QCD like hadron
phase at $N_f=4$. 
The results are reported in these proceedings~\cite{Nagai:lat2011}.

\section{Simulation setup and method}

We adopt the highly improved staggered quark (HISQ) 
action~\cite{Follana:2006rc},
but without the tadpole improvement and 
the mass correction term for heavy quarks.
This action further suppresses the taste breaking in the QCD simulations
compared to the Asqtad or stout-smeared staggered fermions
\cite{Bazavov:2011nk}. 
We use this action for the many flavor simulation  to minimize the
discretization effects 
which potentially compromise the behavior of the physical quantities
at non-zero lattice spacings.

Gauge configurations are generated through HMC algorithm
with various parameter sets for the fermion mass $m_f$, volume and 
the bare coupling $\beta=6/g^2$ for $N_f = 12$ and $16$. 
We measure the mass and decay constant
of the lowest state of the pseudoscalar channel (pion), $m_\pi$ and $f_\pi$,
respectively, which are tested against the chiral symmetry breaking or
conformal scenario. 
The tree-level chiral perturbation theory (ChPT) is used for the former.
The conformal hyperscaling \cite{Miransky:1998dh} and and its
finite-size version is used for the latter.

The finite-size hyperscaling \cite{DelDebbio:2010ze} is derived in the
conformal theory deformed by the small fermion mass and also the large,
finite volume $L^4$. This scaling explains
the fermion mass and volume dependence of the physical quantities,
such as the hadron mass $m_H$, which is
governed by the mass anomalous dimension at the infrared fixed point $\gamma_*$.
According to this scaling, a physical quantity is described by
a function $f(x)$ of the scaling variable
$x = L m_f^{\frac{1}{1+\gamma_*}}$ and $L$,
\begin{equation}
m_H = f(x)/L.
\label{eq:finite_hyper}
\end{equation}
Since the finite-size hyperscaling is an extension of the hyperscaling
to a finite volume,
this scaling should reproduce the original 
hyperscaling~\cite{Miransky:1998dh} in the infinite volume limit,
\begin{equation}
m_H = C_H\, m_f^{\frac{1}{1+\gamma_*}}.
\label{eq:hyper}
\end{equation}
Therefore when the $L$ is large enough,
the function should become
\begin{equation}
f(x) = c_0 + C_H x.
\label{eq:lin}
\end{equation}
We use this fit form in the following analysis, and check if our
data present the scaling behavior or not.

\section{Results of $N_f = 12$}
\label{sec:res_12}

After a pilot study with several bare gauge couplings $\beta \equiv 6/g^2$ 
for the $N_f=12$, we determined to carry out the production run for
$\beta = 3.7$. We generate the gauge ensembles at three volumes,
$L^3\times T$, $(L,T)=(12,24)$, $(18,24)$, $(24,32)$
\footnote{Ideally, the ratio $L/T$ should be fixed, which is not the
case for $L=12$. However, we find the spectrum is not sensitive
to $L/T$ when $m_\pi L\gg 1$, and we only use the data in such a region
in this analysis.},
and the various fermion masses, $0.04 \le m_f \le 0.2$.
We accumulate typically 500--1000 trajectories at each parameter set of
$L$ and $m_f$, and measure the $m_\pi$ and $f_\pi$ on these configurations.

\subsection{ChPT vs.~hyperscaling fit}
\label{sect:ChPT}

\begin{figure}[!h]
\center
\includegraphics*[angle=0,width=0.49\textwidth]{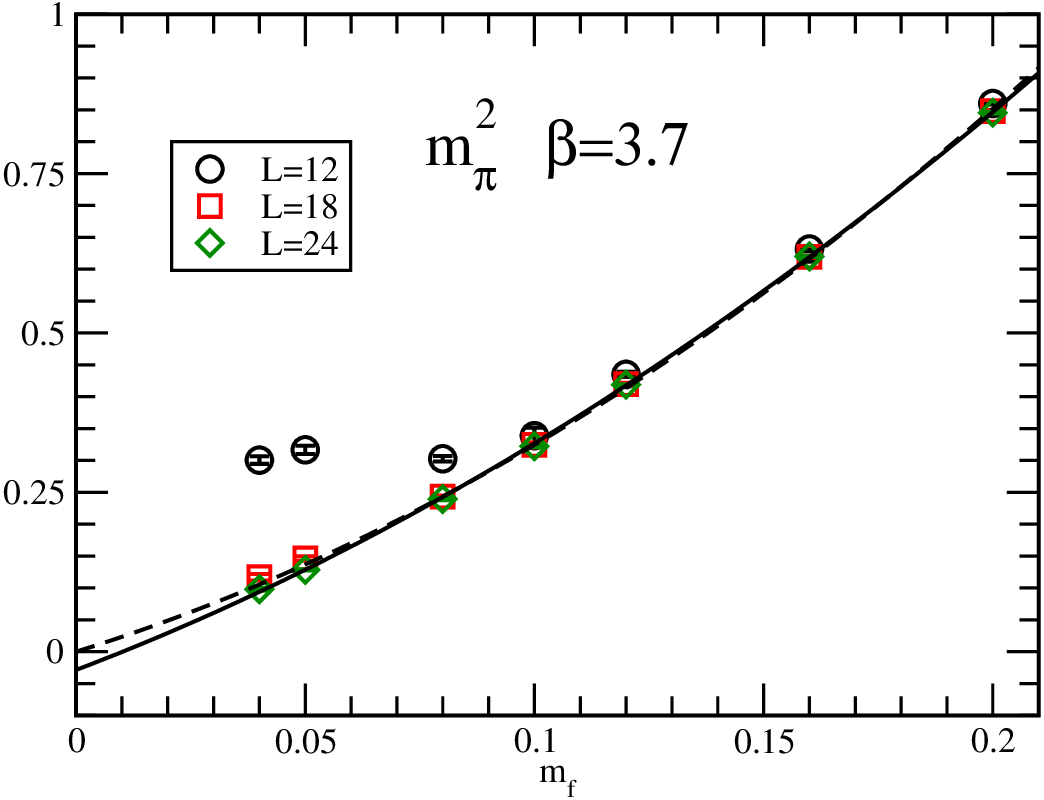}
\caption{$m_\pi^2$ in the 12 flavors as a function 
of the bare fermion mass $m_f$.
The different symbols denote the data at the different volumes.
The solid and dashed lines represent the ChPT fit with and without the 
constant term $c_0$ in eq.~(\protect\ref{eq:chpt}),
respectively, using the largest volume data.
\label{fig:chpt}
}
\end{figure}

Figure~\ref{fig:chpt} shows the result of the $m_\pi$ at each $L$ and 
$m_f$.
In this analysis we use the result obtained from only the largest volume,
$L=24$, because in this volume we expect that finite volume effects 
of the $m_\pi$ is negligible in our fermion mass range.

We first attempt to analyze pion mass with a fit form motivated by 
the chiral perturbation theory (ChPT) given by
\begin{equation}
m_\pi^2 = c_0 + c_1 m_f + c_2 m_f^2.
\label{eq:chpt}
\end{equation}
As the $m_\pi$ should vanish at the chiral limit,
$c_0$ must be zero within the error if the chiral symmetry is broken,
thus, ChPT describes our data.
The values of the $\chi^2/$d.o.f. are 4.1 and 28 for the fit form 
with and without the $c_0$ in eq.~(\ref{eq:chpt}), respectively. 
The former gives better $\chi^2/$d.o.f, while the $c_0$ is non-zero
$c_0 = -0.0287(26)$. 
On the other hand,
the hyperscaling fit eq.~(\ref{eq:hyper}),
gives $\chi^2/\mathrm{d.o.f.} = 3.6$ and $\gamma_* \sim 0.46$.  
This means that the hyperscaling, where the fit gives the smallest
$\chi^2/\mathrm{d.o.f.}$ 
with $m_\pi=0$ at the chiral limit, is favored by our data.

\subsection{Analysis with finite-size hyperscaling}

\begin{figure}[!b]
\hspace{-8mm}
\mbox{
\includegraphics*[angle=0,width=0.27\textwidth]{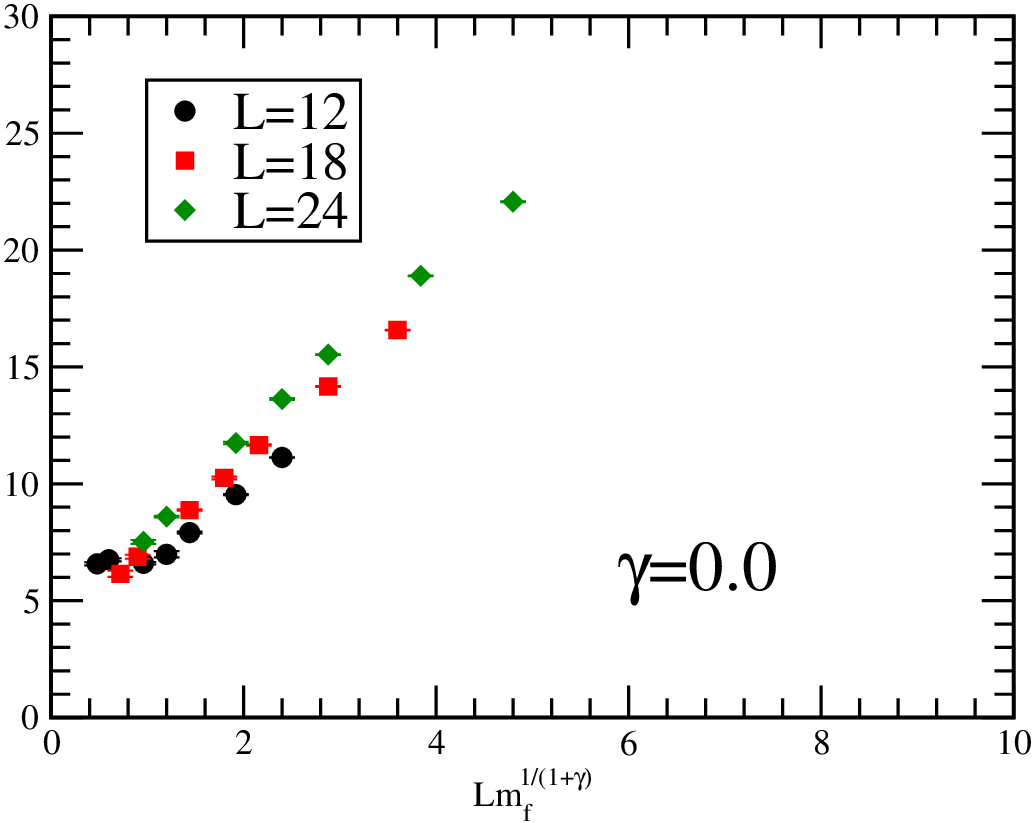}
\includegraphics*[angle=0,width=0.27\textwidth]{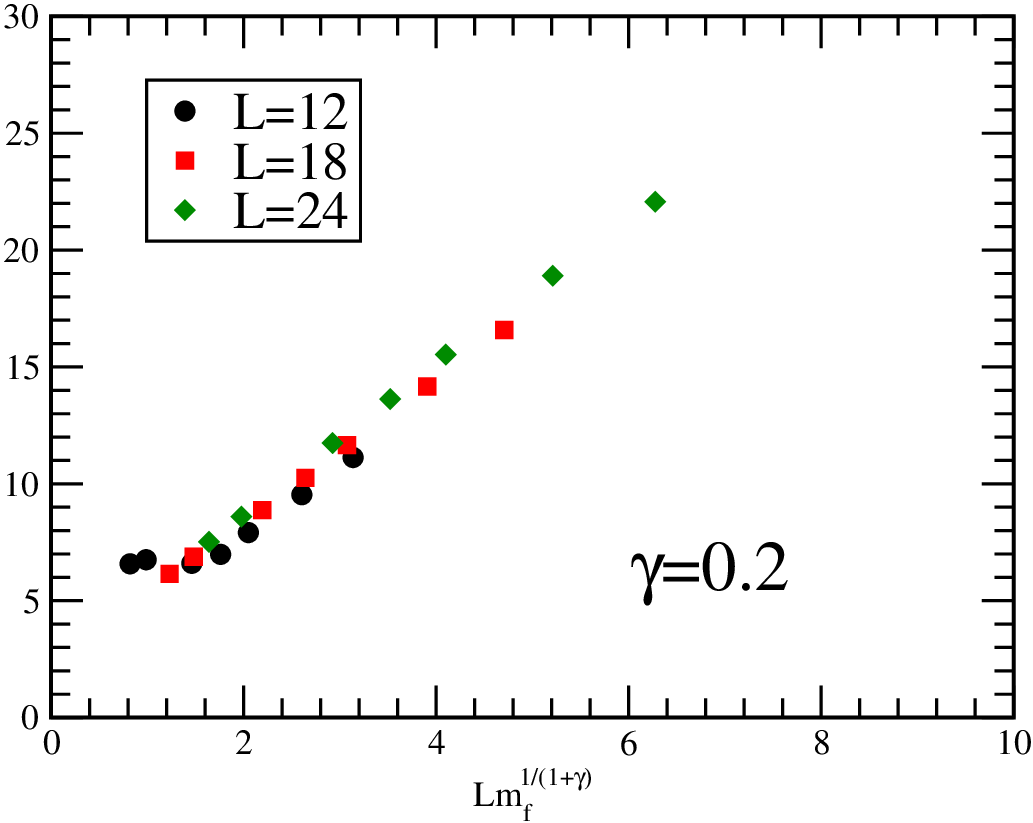}
\includegraphics*[angle=0,width=0.27\textwidth]{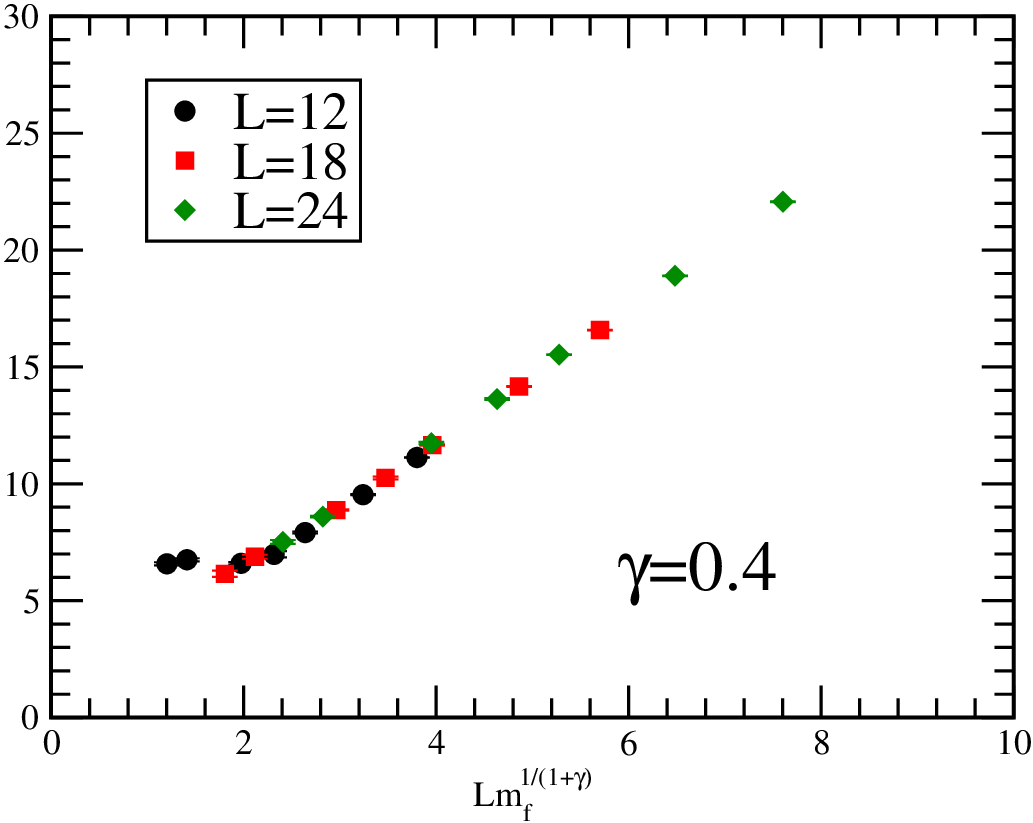}
\includegraphics*[angle=0,width=0.27\textwidth]{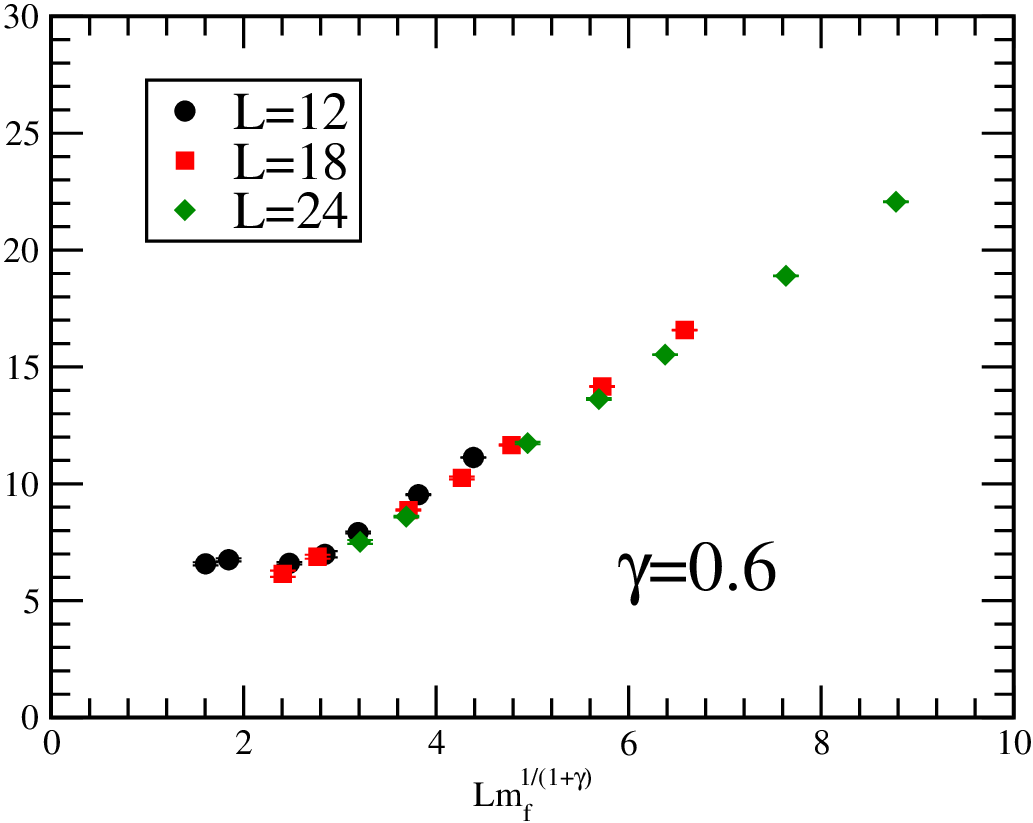}}
\caption{The scaling behavior of $Lm_\pi$ in the 12 flavors as a function of
$x=Lm_f^{\frac{1}{1+\gamma}}$ where the values of $\gamma$ are taken 
as $\gamma=$0, 0.2, 0.4, 0.6 from left to right.
The different symbols denote the data at the different volumes.
}\label{fig:gamma}
\end{figure}

Figure~\ref{fig:gamma} shows the $L m_\pi$ at three lattice sizes
as a function of the scaling variable $x = L m_f^{\frac{1}{1+\gamma}}$,
with $\gamma$ being different in each panel.
In $\gamma = 0$, the data are scattered, but as the $\gamma$
increases, the data in the different volumes tend to align.
The data with $\gamma = 0.4$ show a good alignment,
which is taken as the signal of the scaling, eq.~(\ref{eq:finite_hyper}).
The scaling again disappears beyond $\gamma = 0.4$.
Thus, the optimal $\gamma$ is around $\gamma = 0.4$.
This analysis is based on the method in Ref.~\cite{DelDebbio:2010hu}
for an estimate of the $\gamma_*$.

\begin{figure}[!h]
\center
\includegraphics*[angle=0,width=0.49\textwidth]{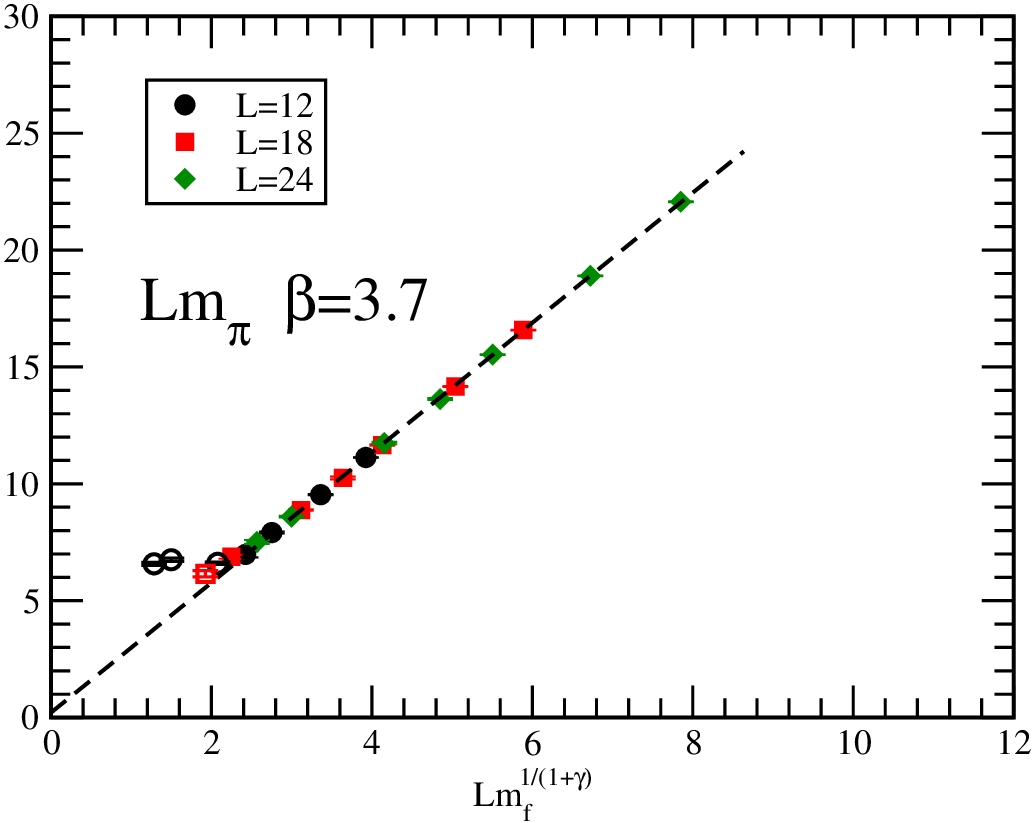}
\caption{
$L m_\pi$ in the 12 flavors and the fit result 
of the finite-size hyperscaling denoted by the dashed line.
The different symbols denote the data at the different volumes.
The filled and open symbols represent the data included in the fit 
and omitted from the fit, respectively.
}\label{fig:fit}
\end{figure}

We attempt to determine the $\gamma_*$ 
by fitting the data to the finite-size hyperscaling with the linear assumption,
eq.~(\ref{eq:lin}).  
Since this assumption is valid in large $L$ region,
we restrict ourselves to use the data in the larger $x$ region.
From the fit shown by the dashed line in Fig.~\ref{fig:fit} we 
obtain $\gamma_* \sim 0.44$ with $\chi^2/\mathrm{d.o.f.} = 4.0$.
The value of $\gamma_*$ is reasonably consistent with the value 
estimated above and also
that obtained from the hyperscaling fit in Sect.~\ref{sect:ChPT}.
We note that our $\gamma_*$ roughly agrees with the result
\cite{Fodor:2011tu} and the results with similar analysis 
\cite{DeGrand:2011cu,Appelquist:2011dp} using the data in
Ref.~\cite{Fodor:2011tu}.

These results are encouraging as preliminary results, while
at present we have not evaluated errors of the $\gamma_*$.
There are possibly several systematic errors in this analysis,
such as the one due to the assumption of the asymptotic form of $f(x)$,
corrections of the finite-size hyperscaling from
large $m_f$ and small $L$ effects~\cite{Aoki:2012ve}.
If the effect of these corrections is large, we cannot obtain
the correct $\gamma_*$ even if the data shows the scaling behavior.
In order to get reliable estimate of the systematic errors, we need to expand
our simulation towards lighter mass and larger volume.
Detailed analysis using those data are ongoing.

\section{Results of $N_f = 16$}
\label{sec:res_16}

In the $N_f = 16$ SU(3) gauge theory we perform the simulations
at $\beta \equiv 6/g^2 = 3.15$ and 3.5.
We choose three spatial volumes $L=8, 12, 16$ at $\beta = 3.15$, 
and add one more volume $L = 24$ at $\beta = 3.5$.
The range of the fermion mass is $0.03 \le m_f \le 0.2$.
The typical length of the trajectory is roughly 1000.

We analyzed the results of the $m_\pi$ and $f_\pi$ in the $N_f = 16$ case
using the same procedure as in the $N_f = 12$ case discussed in the previous
section.
The ChPT fit failed as it did for $N_f=12$,
which is expected as the $N_f = 16$ theory is much deeper in conformal
window if $N_f=12$ was.

\begin{figure}[!t]
\includegraphics*[angle=0,width=0.49\textwidth]{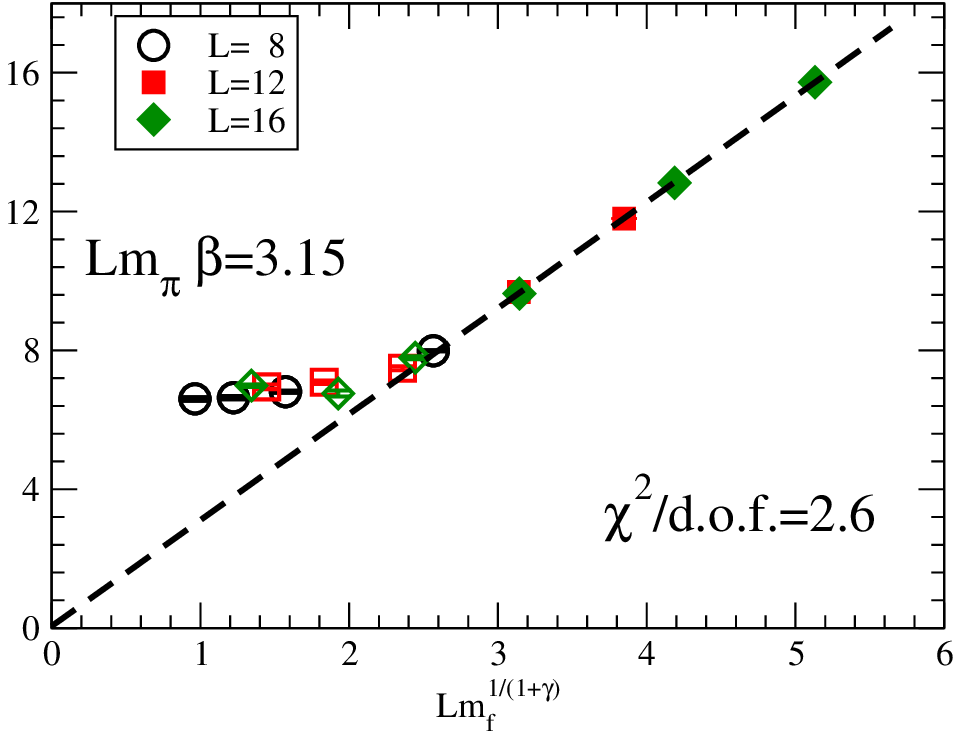}
\includegraphics*[angle=0,width=0.49\textwidth]{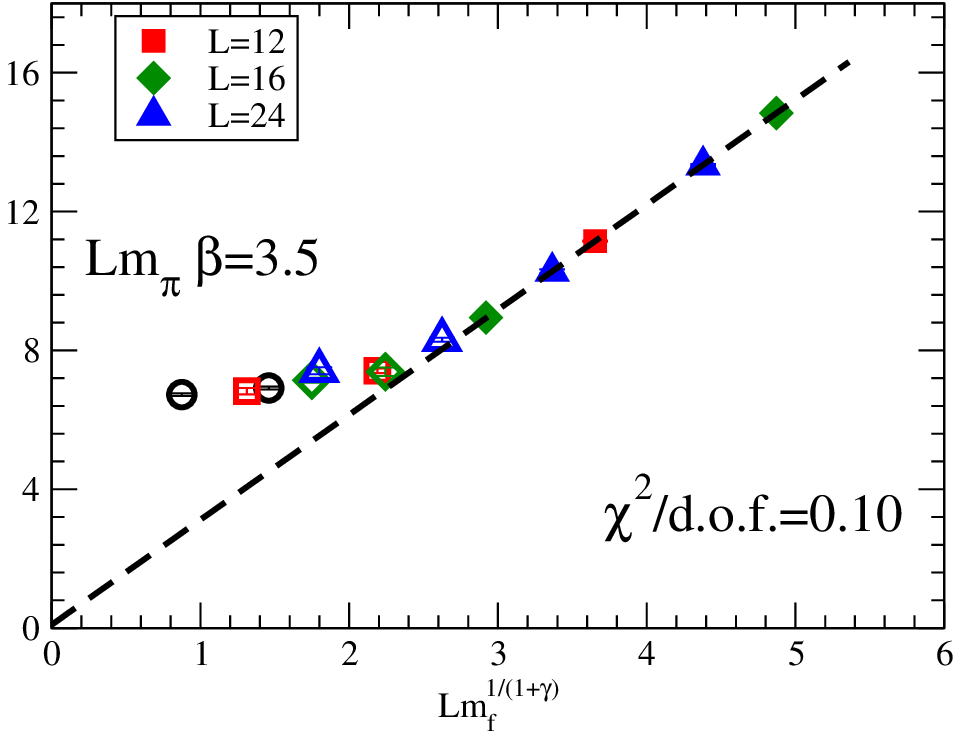}
\caption{
$L m_\pi$ in the 16 flavors at $\beta = 3.15$(left) and $\beta = 3.5$(right).
The different symbols denote the data at the different volumes.
The dashed line denotes the fit result of the finite-size hyperscaling.
The filled and open symbols represent the data included in the fit 
and omitted from the fit, respectively.
\label{fig:4}
}
\end{figure}

Using the finite-size hyperscaling analysis,
we attempt to determine the value of the $\gamma_*$ at two $\beta$'s.
Figure~\ref{fig:4} shows the fit result of the $L m_\pi$
with the finite-size hyperscaling assuming the 
asymptotic form of $f(x)$, eq.~(\ref{eq:lin}).
The both panels show good scaling behavior of the $L m_\pi$,
and the sizes of the $\chi^2/$d.o.f. are reasonable.
From the fits we obtain $\gamma_* \sim 0.42$ and 0.35 at
$\beta = 3.15$ and 3.5, respectively.
We also analyze the data of $f_\pi$ and obtain $\gamma_*$
which is roughly consistent with those with $m_\pi$.
Again we have not yet estimated errors of the obtained $\gamma_*$
discussed in the above section.

These observations lead to a converged picture of the conformal theory
with decreasing $\gamma_*$ towards weaker bare coupling.
One question arises as to the size of the $\gamma_*$ being much
bigger than the result with 2-loop perturbation theory 
$\gamma_*^{\rm pert} \sim 0.025$.
One possible scenario is that the $\gamma_*$ further decreases 
towards much weaker coupling (continuum limit) and eventually
gets compatible to the perturbation theory.
Another possible scenario is our bare gauge coupling is too 
large to investigate the property of this theory in the continuum limit.
To investigate further, more detailed study of 
the $\beta$ dependence of the $\gamma_*$ is necessary.

\section{Summary and outlook}
\label{sec:summary}

We have studied the SU(3) gauge theories with the fundamental 12 and 16
fermions using a HISQ type staggered fermion action, and presented
preliminary results in this report.
For the 12-flavor case, we attempt to determine the phase of this theory
through the analysis of pion mass.
Our present data favors the conformal hyperscaling over the ChPT.
The mass anomalous dimension, $\gamma_*$ at the infrared fixed point
was estimated though the (finite-size) hyperscaling analysis. 
So far the size of $\gamma_*$ is not as big as $\gamma_*\sim 1$
for the theory to be close to the realistic technicolor theory.
We have not yet estimated the errors of the $\gamma_*$,
so that we need to estimate it to reach a definite conclusion.
More detailed analyses with more data at larger volume and lighter mass
are underway.
For the 16-flavor case, our data shows the conformal signal, while
the obtained value of the $\gamma$ is much higher than the perturbative
result. We plan to investigate this by studying the $\beta$ dependence
of this value.

\section*{Acknowledgments}
Numerical calculations for the present work have been carried out
on the cluster system ``$\varphi$'' at KMI, Nagoya University with
the public code provided by the MILC Collaboration.
We thank Katsuya Hasebe for the useful discussion.
This work is supported in part by Grants-in-Aid for Scientific Research
from the Ministry of Education, Culture, Sports, Science and Technology 
(Nos. 
22224003, 
23540300, 
21540289
) and 
Grants-in-Aid of the Japanese Ministry for Scientific Research on Innovative 
Areas (No. 
23105708
).

\bibliography{kmi_nf12+16}

\end{document}